# Chaotic states of rf biased Josephson fluxonic diode

Hamed Mehrara, Alireza Erfanian, Farshid Raissi

The results of chaos simulations in Josephson Fluxonic Diode (JFD) subject to the magnetic field, rf driving and dc biasing currents in the RCSJ model are reported. The dynamic behavior of this nonlinear device is mapped as a function of loss, bias, frequency and amplitude of rf excitation which leads to find sustainable fluxon oscillation regions in microwave application. As a result, chaos does not happen if either the dc bias or the rf excitation is below specific value, but the loss term can change this state without affected by rf frequency. An attempt is made to explain the boundaries of the chaotic region in terms of parameter maps.

**INTRODUCTION**

One of the most fascinating features of nonlinear physical systems is that they can exhibit spatial and temporal structures, which can be described both by their coherent and chaotic natures [1]. In superconducting electronics and digital communication, such chaotic solutions have a psuedo-random nature which yields a broadband component in the power spectrum like that of a random noise source [2-4]. On the other hand, the existence of chaos is of importance to practical devices such as parametric amplifiers voltage standards [5,6] and pulse generators [7].

Simulation results of chaotic behavior in Long Josephson Junctions (LJJs) modeled by the perturbed sine-Gordon equation in constant magnetic fields showed that dc current bias cannot induce chaotic flux dynamics, while AC bias current can do [8-11]. In the parameter space, there is no clear regular boundary between local chaotic attractors and local regular attractors. But, they consist of period-doubling bifurcation, intermittency between and within Fiske steps [12,13]. This result demonstrates that for such multidimensional systems chaos is really low dimensional and in specific regions quasiperiodic motion can be found [14]. Moreover, phenomena such as creation or annihilation of fluxons, which do occur in real LJJs, are not considered and in order to avoid the occurrence of these phenomena, the amount of energy which is injected into the system kept small. This is why the chaos analysis in LJJs is restricted to small rf field amplitudes.

In this paper Josephson fluxonic diode as dual of semiconductor p-n junction in superconducting electronics, is considered for its periodic and chaotic states at the presence of rf excitation. Periodic fluxon motion can be generated when a Josephson fluxonic diode is driven by a uniform dc bias with an external spatially reversing magnetic field called control current [15]. This phenomenon is evidenced by singularities in the current-voltage characteristic of JFD in its forward and reverse bias. It provides an asymmetric IV curve in which the forward bias corresponds to a voltage state and reverse bias to a short circuit [16]. Modeling of this nonlinear system starts from the perturbed sine-Gordon equation describing the behavior of a Josephson junction as one unit cell of JFD combined with applied dc and rf currents:

$$\varphi_{tt} - \varphi_{xx} + \frac{1}{\sqrt{\beta}}\varphi_t + \sin\varphi = \rho_b + \rho_c + \rho_\omega \sin\Omega t \quad (1)$$

In this equation, the $\varphi(x,t)$ is phase difference of JFD's electrodes and has soliton-like solutions which correspond to magnetic field quanta inside the JFD, normally called fluxon or vortices and dependents on McCumber hysteresis parameter ($\beta$), frequency ($\Omega$), amplitude of radiation ($\rho_\omega$) and finally applied dc control and bias current ($\rho_b+\rho_c$). For simplicity, all length parameters are measured in units of the Josephson penetration depth $\lambda_J$, time in units of $1/\omega_J$ where $\omega_J$ is the Josephson plasma frequency and current amplitudes by maximum Josephson supercurrent $I_c$. Typically, the Resistively and Capacitively Shunted Josephson Junction (RCSJ) model is found an appropriate approach to study $\varphi(x,t)$ solutions for chaos examination in long Josephson junction which can be used for JFD as well [16, 17]. Although, deterministic results for rf-biased Josephson junction expresses partially chaotic behavior [18-20] but in this study we have motivated to investigate whether JFD with its special vortex dynamics, can act periodically or chaotically in digital communication systems. We report the first systematic study which attempts to define the conditions for which a coherent operation of JFD is obtained. The structure of this paper is as follows: In Section 2, we provide all the necessary preliminaries including the basics on Josephson fluxonic diode, the RCSJ modeling, chaos analyzes methods including Poincare map, bifurcation diagrams and power spectra which are described in continue. Section 3 contains the results and the discussion, and finally, Section 4 summarizes the main conclusions for this work.

## THEORETICAL BACKGROUND

JFD is a long Josephson junction to which a spatially reversing magnetic field is applied [15]. Vortices fill half of the junction and anti-vortices fill the other half as the phase carriers in Josephson junction medium. Control current value determines the number of these vortices in a JFD. By appling a uniform bias current in the junctions, solitons and antisolitons push (pull) to (from) each other using a Lorentz force which its lateral direction depend on bias polarity. The distance at the center of the JFD, in which the polarity of the magnetic field changes, is termed transition region. Fig. 1 demonstrates plane configuration of a JFD in which a separate control line, running above the base electrode, creates an asymmetric magnetic field at boundaries and has a rectifying type of current–voltage curve.

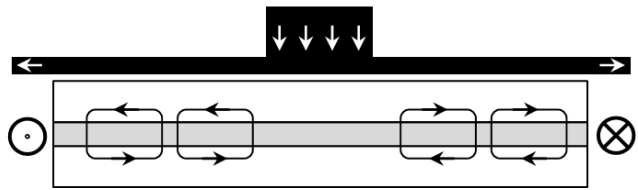

Fig. 1. Configuration for a JFD: a separate control line (black colored) running above the base electrode of a long Josephson junction (white/gray colored) creates a spatially reversing magnetic field inside and generates vortices and anti-vortices in each side (dimension is not to scale).

In this limit, the generated fluxon and anti-fluxon pairs by using dc control current is uniformly distributed in the junction. However, compared to a long Josephson junction, JFD provide distinct advantages. It has inherent gain, can determine the frequency and amplitude of radiation simultaneously and needs very simple addressing scheme and its size can be varied providing a much simpler circuit [21]. Chaotic behavior described in Eq. 1 represents the ideal behavior of a unit cell in JFD. However, the characteristics of JFD can accurately estimates according to an equivalent electrical model; namely RCSJ in which each Josephson junction unit cell is described by an ideal junction J shunted by resistance $R_n$ and capacitance C to form a parallel circuit. Series elements in this model rise from JFD superconductor's plate. The surface impedance of a superconductor plate can be expressed by a complex number whose imaginary part is positive. The real part of this impedance represents the electric loss which manifests itself as a series resistance $R_c$ in the line and its positive imaginary part shows that superconductor plate is inductive by L. The detailed study of chaos in JFD depends on the parameter values and on the boundary conditions, which will be introduced in the following for the particular conditions. When the applied rf excitation is symmetric on at both side, the boundary conditions for the first and last unit cell of JFD will be:

$$\varphi_x |_{x=L/2} = \rho_b + \rho_c + \rho_\omega \sin \Omega t$$
$$\varphi_x |_{x=-L/2} = \rho_b - \rho_c + \rho_\omega \sin \Omega t \qquad (2)$$

where *L* is total length of JFD. In order to investigate how fluxon oscillations are affected by an rf bias and other parameters, circuit simulation were performed in which the steady-state behavior was determined for a selection of rf amplitudes and frequencies in the range $0.01 < \rho_\omega < 0.1$ and $0.1 < \Omega < 1$ respectively. At the beginning, the rest condition ($\rho b = \rho_c = 0$) is considered. Then control current is fixed to $\rho_c = 0.15$ with $-1.0 < \rho_b < 1.0$ and $\beta$ varies to be 10,100 and 1000. For these values of dc bias, the JFD supports fluxon oscillations on different operating points and includes a frequency to be called the proper fluxon oscillation frequency. The algorithm includes totally 1000 unit cell for $L=70\lambda_J$, $J_c=700$ A/cm$^2$ and $\omega_J =140$ GHz at voltages below the energy gap. Also, the transition region is about $20\lambda_J$. In all cases the rf steady-state behavior is characterized by applying dc control and bias initially, then wait for a steady-state response and in continue, rf excitation will apply. So we inspect the solution over at least 10 rf periods, but in cases where chaos is present at least 50 rf periods will be computed and often computations extends to 100 rf periods.

In practical chaotic circuit exploration, it is useful to have a detailed knowledge about the dynamical behavior of the system in different parameter ranges, so that one can avoid certain undesirable regions in the parameter space to optimize the performance. For that purpose, tools include in this study are: Poincare map in the phase space, and power spectra [22]. The Poincare map in phase space was displayed by having the $\varphi_t$ of simulation results (which is proportional to output voltage) and map the position in phase space for every rf cycle. In this way the order of a sub-harmonic oscillation was determined as the number of points in the Poincare map and chaos appeared typically as a strange attractor. More accurately, power spectra allow a non-periodic signal to be decomposed into harmonic signals by using spectrogram. The power spectrum of a periodic signal presents a sharp peak at the rf excitation frequency $\Omega$ and its harmonics, as well. Alternatively, a quasi-periodic signal will show several frequency peaks as well as their lineal combinations. However, a chaotic signal is characterized by a continuous power spectrum like that of a random noise source along the frequency.

**RESULTS AND DISCUSSION**

The solutions of the RCSJ model for JFD have been computed for different values of the parameters and initial conditions. It is worth pointing out that the simulations have been carried out via MATLAB. It contains an implementation with time-dependent simulation. Primarily, we compare the results for an ideal case, by choosing the rest parameters $\rho_b = \rho_c = 0$ and limit $\Omega$ to 0.1 for very low rf amplitude ($\rho_\omega = 0.01$). In this case, we have considered the $\beta = 10$. Fig. 2.a shows the trajectory of the normalized average phase difference $\varphi_t$ vs. $\varphi$ along the length of JFD and Fig. 2.b displays spectrogram of $\varphi(x,t)$ for each unit cell of JFD. Spectrograms are useful in that they clearly indicate the components of a periodic response and show chaotic regions allow plotting of multi-parameter, especially two-parameter, systems [23]. The ordinate of the spectrograms is the response frequency; while the contour indicates the power of the system (dark colors are low power, light colors high power) at that frequency. Such a graphical representation shows normalized peak of power spectrum by a color for each cell in x direction and generated harmonic frequencies along y- axis which are normalized to $\omega_J$ and shifted zero-frequency component to center of spectrum. To track broadband chaotic state, harmonic frequency illustrated up to $50\omega_J$. These frequencies correspond to the frequency of the phase difference in the stationary state.

Base on the results, in non-chaotic Phase space plots (Fig 2.a) the trajectory resembles an elliptical orbit repeating itself continuously, since it returns to the same point under the same conditions. On the other hand, Fig .2.b manifests the same information and shows one main peak at the middle of power spectrum plane. This white-colored peak fades horizontally which means that at the center of simulated JFD there is some variation in phase due to rf excitation at boundaries. Under this circumstance, the presence of one clear and sharp frequency peak constitutes a strong indicator of chaos absence. It is worth noting that in this state, the same non-chaotic results will be acquired by $\beta =100$ and 1000 or increase $\Omega$ to 1 which are not presented here.

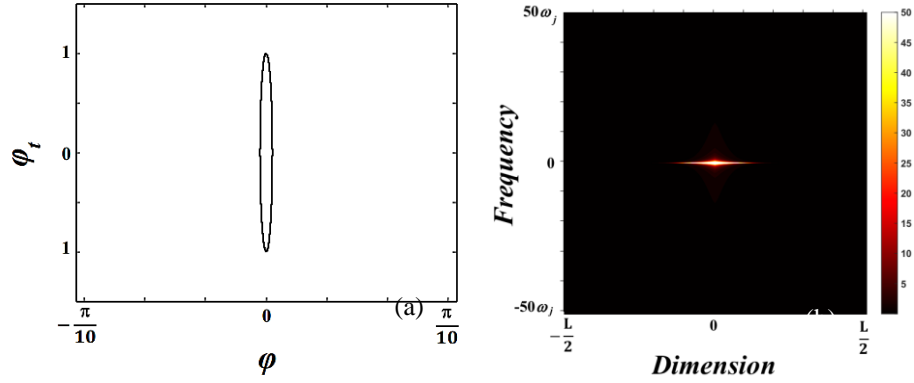

Fig. 2. Chaos check along the length of JFD for rest parameter values ($\rho b=\rho_c=0$) and $\rho_\omega = 0.01$, $\Omega =0.1$ and $\beta = 10$, (a) The normalized average phase variation ($\varphi_t$) vs. phase ($\varphi$). (b) spectrogram of $\varphi(x,t)$ for each unit cell of JFD vs. produced harmonic frequencies. In non-chaotic Phase space plots, the trajectory resembles an elliptical orbit repeating itself and spectrogram presents one clear and sharp frequency peak along horizontal axis at the center (Dark colors are low power and light colors high power.)

In this situation, increasing the amplitude of rf excitation up to a threshold ($\rho_{th}$), will change periodic state to chaotic. Fig. 3 shows phase space and spectrum for parameters choice: $\Omega =0.1$ and different $\beta$ in rest condition. $\rho_\omega$ is selected slightly below $\rho_{th}$ at the edge of periodic to chaotic transition. In these cases, the phase space displays an orbit changing. Thus, the elliptical path stays periodic but repeats itself with a new period, having two orbits or more instead of only one. This phenomenon occurs in nonlinear systems by amplitude increasing, and consists of a bifurcation of the original loop as the number of time required to return to the original state changes. The $\rho_{th}$ for this bifurcation decrease as $\beta$ increase (0.0895, 0.0855 and 0.0850 respectively for $\beta$=10 (Fig 3.a, b), 100 (Fig 3.c, d) and 1000 (Fig 3.e, f)). The implication is that the margin of chaotic state shrinks with $\beta$ growth and in spectrum, presence of sharp and clear central frequency peaks confirms un-chaotic behavior; however a spatially appearing periodic peak in spectrogram resonates with the number of orbits in phase diagram. There can be two, three or more bifurcations in an infinite sequence of orbit changing as $\rho_\omega$ increases. For higher values of $\rho_\omega$, the number of bifurcations grows and If the number of bifurcations is infinite, then the system behaves chaotically like what has been done for Fig. 4. In this case, the parameters have been chosen to be $\Omega$=0.1, $\beta$=10 and $\rho_\omega$=0.09 in rest condition.

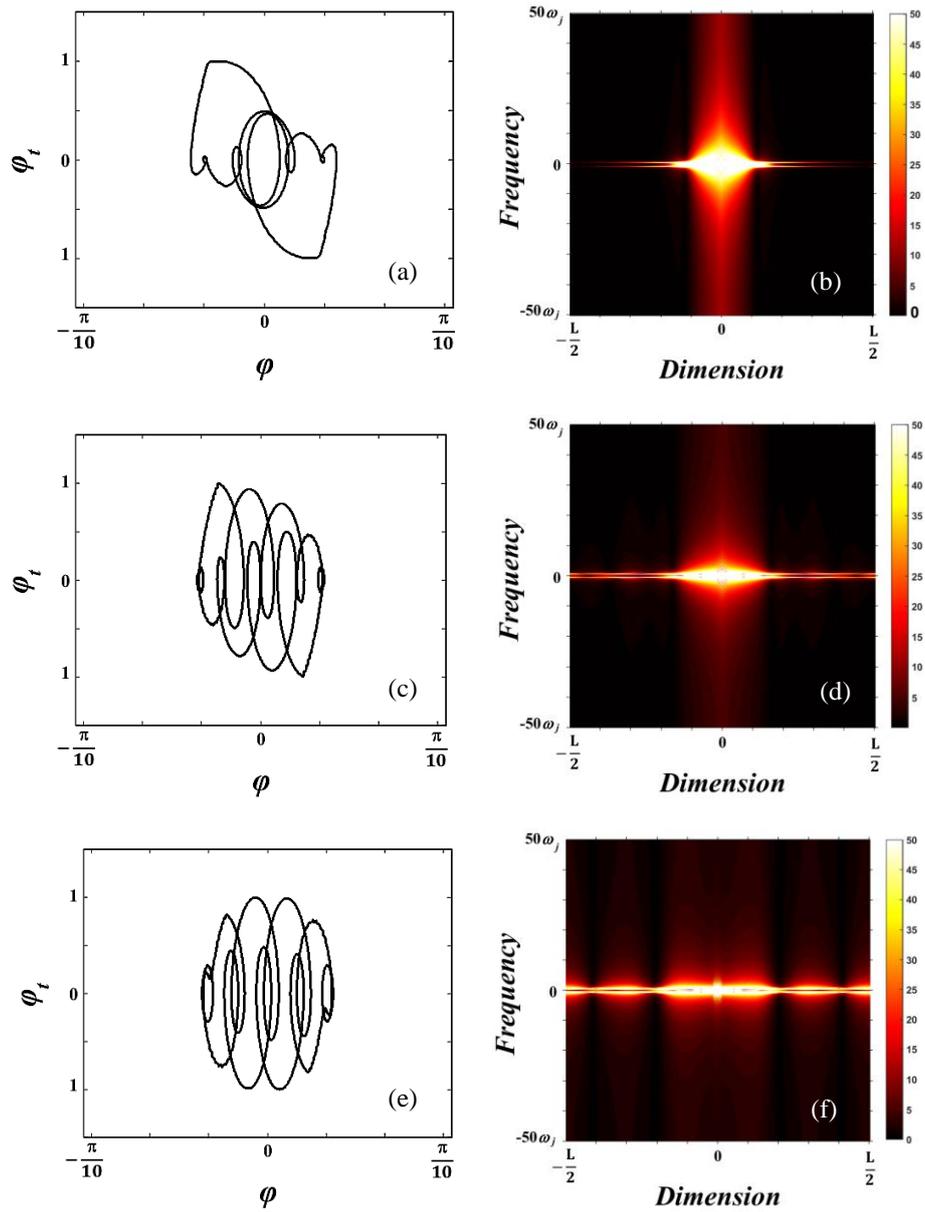

Figure 3. Phase variation and power spectrum under the same parameters as in Fig 2. Other parameter values are chosen to be $\Omega =0.1$ for (a, b) $\beta=10$ (c,d) $\beta=100$ and (e,f) $\beta=1000$. The trajectory shows orbit changing, since it does not return to the initial conditions until it completes some orbits.

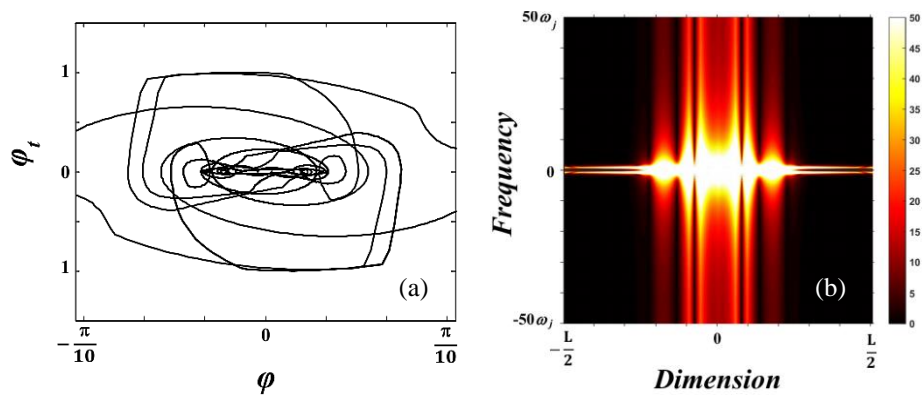

Figure 4. A Chaotic trajectory in: (a) phase diagram and (b) spectrogram of the trajectory. In this case, the parameters have been chosen to be $\rho b=\rho_c=0$, $\Omega =0.1$, $\beta = 10$ and $\rho_\omega = 0.09$.

On the other hand, if the rest parameters change to a biased one ($\rho_b \neq 0$, $\rho_c \neq 0$), JFD will forward biase to a voltage state or reverse to a short circuit mode. Principally, application of the bias sets solitons and antisolitons in motion by a Lorentz-type force and pushes or pulls them based on bias polarization. Depending on the level of loss, they remain stationary, pass through each other or annihilate with each other. In a low loss junction with $\beta=1000$, they move and pass through each other while in a lossy junction of $\beta=10$ the annihilating will happen [17]. These two different phenomena are prone to generate chaos in which for our study the annihilating has been considered. So, let us choose control current to be $\rho_c = 0.15$ and varies $\rho_b$ in forward and reverse bias, then chaotic zone can be extracted, leading to map diagram provided in Fig. 5. The main result of our calculations is this map which can be divided into two parts: the blue colored area labeled "Periodic" in which the solutions are periodic, and a red one labeled "Chaotic" in which the solutions are frequently chaotic. The colors are related to the number of harmonic peaks as explained in Ref [23] in frequency response of JFD and conditions are same as for Fig. 2. As depicted in Fig. 5, the boundaries between the chaotic and periodic regions shown here is sharp to distinct these two circumstances. In addition to these two solutions, other regions include chaotic transients, simply periodic, and complicated periodic responses are also observed which are colored between blue and red.

In Fig 5.a, no fluxon movement is produced without a bias, thus increasing rf excitation only vibrates the stationary solitons and antisolitons by frequency and consequently phase will be periodic up to a specific excitation ($\rho_{th}$) which depends on loss parameter. At a higher bias value and for this low loss case, the annihilating starts and changes state to chaotic one. There are some significant points in Fig 6.a with low rf excitation in which the annihilating does not violate average phase periodicity. These points are seen in forward bias but not in reverse (Fig 6.b) which means vibrating and moving solitons and antisolitons annihilate without perturbation and the consequent released energy is consumed by transient vibrating in sub-division of solitons and antisolitons. In reverse bias, the minimum of $\rho_b$ for jumping to chaotic state is more than forward bias and points out that reverse bias persists chaos generation by pulling solitons and antisolitons away while the forward bias have no obligatory persistence.

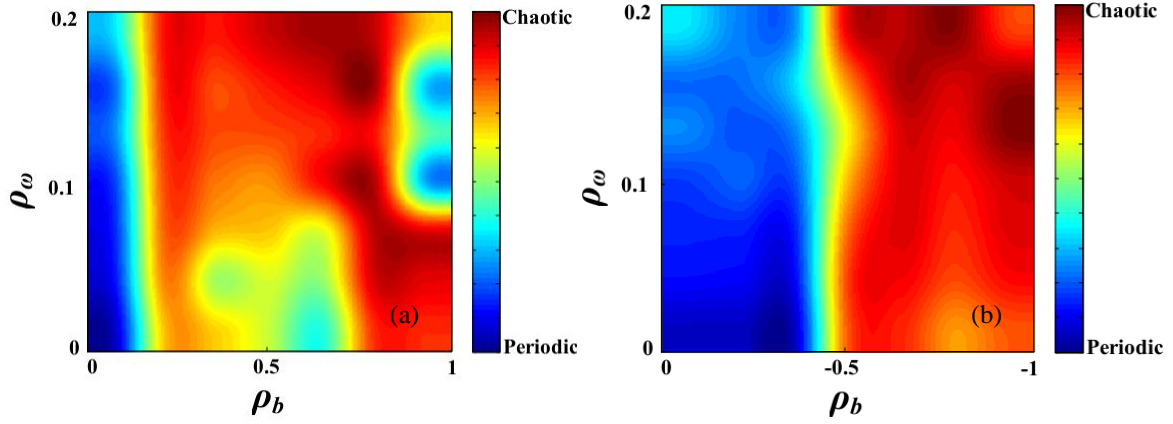

Fig. 5. Chaos map in the ($\rho_\omega$, $\rho_b$) plane with $\Omega = 1$, $\rho_c = 0.15$ and $\beta = 10$, for: (a) forward biased and (b) reverse biased JFD.

The second important map is indicated in Figure. 6 a and b. Fig. 6.a encircles the region of the ($\rho_b$, $\beta$) plane in which the steady-state solutions are spatially uniform in the presence of the rf drive even though fluxon oscillations existed before the rf was turned on. It depicts an interesting phenomenon regarding the chaotic behavior of the JFD that increase of both $\rho_b$ and $\beta$ presents irregular oscillations with orbit changing connected with chaotic jumps. Hence, spectrogram power of lossless JFD, almost produce noise-like spectrum in this case. At any rate, it presents two peaks, located at $\rho_b = 0.45$ and $\rho_b = 0.55$, respectively in which are quite noisy and the former can be suppressed by choosing a lossy JFD. Nonetheless, the dynamics in Fig. 6.b appears that increasing the frequency of rf excitation in a biased JFD will not change the state of periodic phase considerably (if we neglect the changes in characteristic parameters of Josephson junction) whereas $\beta$ which is a fabrication related parameter can generate chaos.

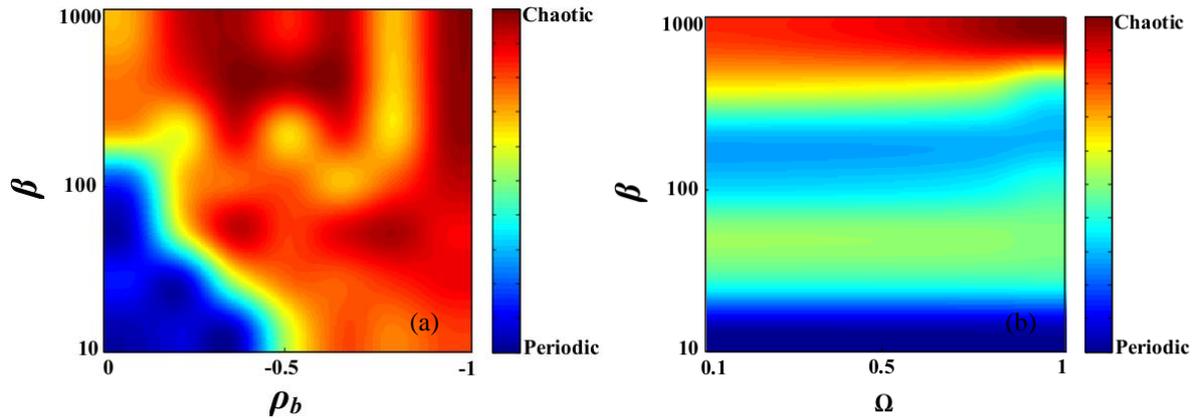

Fig. 6. Map of the chaos for $\rho_c = 0.15$ and $\rho_\omega = 0.1$, in: (a) the ($\rho_b$, $\beta$) plane with $\Omega = 1$, (b) the ($\Omega$, $\beta$) plane with $\rho_b = -0.3$.

## CONCLUSIONS

In summary, we have investigated chaotic dynamics of a Josephson fluxonic diode (JFD) by using RCSJ model for a specific set of parameters in a wide range to obtain chaotic and periodic behaviors in JFD. We exploit phase diagrams and power spectrum which show that at the rest condition with no bias and control current, amplitude of rf signal is able to change phase state from periodic to chaotic in a lossless JFD while the frequency of excitation can do the same. On the other hand, under simultaneous application of control and bias current, annihilation appose in forward biased which in a lossy JFD exhibit periodic behavior and increasing $\beta$ will jump this state to chaotic ones. In reverse bias, the same behavior happens however the periodic region is much extensive than forward bias that indicates the possibility to design un-chaotic oscillators with considerably great energy by using the method for controlling chaos.